\documentclass[aps,preprint,prd,showpacs,nofootinbib]{revtex4}

\usepackage{latexsym}
\usepackage{amsmath}
\usepackage{graphicx}
\usepackage{subfigure}
\usepackage{dcolumn}
\usepackage{bm}
\usepackage{amssymb}
\usepackage{latexsym}
\usepackage{ulem}
\usepackage{color}
\usepackage[colorlinks,linkcolor=magenta,anchorcolor=cyan,citecolor=blue]{hyperref}

\def\be{\begin{equation}}
\def\ee{\end{equation}}
\def\ba{\begin{eqnarray}}
\def\ea{\end{eqnarray}}

\def\nn{\nonumber}
\def\lf{\left}
\def\rt{\right}

\begin{document}

\title{Higher order derivative coupling to gravity and its cosmological implications }

\author{Yong Cai$^{1}$\footnote{caiyong13@mails.ucas.ac.cn}}
\author{Yun-Song Piao$^{1,2}$\footnote{yspiao@ucas.ac.cn}}

\affiliation{$^1$ School of Physics, University of Chinese Academy
of Sciences, Beijing 100049, China}

\affiliation{$^2$ Institute of Theoretical Physics, Chinese
Academy of Sciences, P.O. Box 2735, Beijing 100190, China}

\begin{abstract}

We show that the $R^{(3)}\delta K$ operator in effective field
theory is significant for  avoiding the instability of
nonsingular bounce,  where $R^{(3)}$ and $K_{\mu\nu}$ are the
three-dimensional Ricci scalar and the extrinsic curvature on the
spacelike hypersurface, respectively. We point out that the
covariant Lagrangian of $R^{(3)}\delta K$, i.e., $L_{R^{(3)}\delta K}$, has the second order
derivative couplings of scalar field to gravity which do not
appear in Horndeski theory or its extensions, but does not
bring the Ostrogradski ghost.  We also discuss the possible
effect of $L_{R^{(3)}\delta K}$ on the primordial scalar
perturbation in inflation scenario.


\end{abstract}

\maketitle

\section{Introduction}

Recently, the studies of the origin of  the universe and the current
accelerated expansion have greatly promoted the development of
gravity theories beyond general relativity (GR), see
\cite{Rubakov:2014jja}\cite{Joyce:2014kja}\cite{Nojiri:2017ncd}
for recent reviews. How to design a theory without extra degree of
freedom (DOF) has acquired persistent attention.

The Horndeski theory was proposed in the 1970s
\cite{Horndeski:1974wa}, see also
\cite{Deffayet:2011gz}\cite{Kobayashi:2011nu}, in which the
equations of motion have at most second order time derivatives,
which avoids the extra DOF (so the Ostrogradski ghost). However,
it seems that equations of motion with higher order time derivatives  do not
necessarily suggest the presence of extra DOF. The discoveries of
the beyond Horndeski theory
\cite{Zumalacarregui:2013pma}\cite{Gleyzes:2014dya}\cite{Gleyzes:2014qga}
and degenerate higher order scalar tensor (DHOST) theory
\cite{Langlois:2015cwa}\cite{Langlois:2015skt}\cite{BenAchour:2016fzp}\cite{Langlois:2017mxy}
have confirmed this possibility and greatly enriched our
understanding of gravity. In Horndeski and DHOST theories,
the Lagrangian involves only the nonminimal couplings $f(\phi,X)R$
and $\phi^{\mu\nu} R_{\mu\nu}$, where $X=\phi_\mu\phi^\mu$, $\phi_\mu=\nabla_\mu\phi$ and $\phi^{\mu\nu}=\nabla^\mu\nabla^\nu\phi$.

Along a different line, the effective field theory (EFT) of
cosmological perturbations has been developed for investigating
inflation \cite{Cheung:2007st}\cite{Weinberg:2008hq} and current
cosmological acceleration
\cite{Gubitosi:2012hu}\cite{Gleyzes:2013ooa}\cite{Piazza:2013coa},
see \cite{Naskar:2017ekm} for a review. Recently, the EFT has
also been  applied to the nonsingular cosmologies
\cite{Cai:2016thi}\cite{Creminelli:2016zwa}\cite{Cai:2017tku}. It
was found first in Refs.
\cite{Cai:2016thi}\cite{Creminelli:2016zwa} that the operators
with three-dimensional Ricci scalar $R^{(3)}$, especially
$R^{(3)}\delta g^{00}$, could play a significant role in curing the
gradient instability induced by a negative sound speed
squared (i.e., $c_s^2<0$) of scalar perturbation
\cite{Libanov:2016kfc}\cite{Kobayashi:2016xpl}. Actually, as will
be shown, the operator $R^{(3)}\delta K$ ($K$ is the extrinsic
curvature) could play a role similar to that of $R^{(3)}\delta
g^{00}$.

We built a fully stable cosmological bounce scenario in Ref.
\cite{Cai:2017dyi} by applying a least set of operators ($(\delta
g^{00})^2$ and $R^{(3)}\delta g^{00}$), namely, a ``least
modification". The graviton throughout the bounce behaves itself like
that in GR, which could naturally avoid the strong
coupling regime appearing in \cite{Ijjas:2016vtq}, see
also \cite{Yoshida:2017swb}. The covariant Lagrangian proposed in
\cite{Cai:2017dyi} belongs to beyond Horndeski theory, (see also
\cite{Kolevatov:2017voe} for a different implementation of a fully stable
bounce), which is a subclass of the DHOST theory, but the
equations of motion still could be second order in time
derivatives. This enlightens us that there might still be
some space of scalar-tensor theory to be explored.

As will be pointed out, the covariant description of
$R^{(3)}\delta K$ contains the second order derivative 
couplings of the field $\phi$ to gravity, such as
${\phi^\mu\phi_{\mu\nu}\phi^\nu}R$, 
$\phi^\mu\phi^\nu(\Box\phi)R_{\mu\nu}$ and
${\phi^\mu\phi^\nu\phi_\rho\phi^{\rho\sigma}\phi_\sigma}R_{\mu\nu}$,
which do not appear in Horndeski (or even DHOST) theory.
The mimetic gravity with the  coupling $(\Box\phi)R$ has
been proposed in Ref. \cite{Zheng:2017qfs}. In scalar-tensor
theory, it is interesting to explore the possibility of such
higher order derivative couplings.


In this paper,
we point out that the covariant Lagrangian of
$R^{(3)}\delta K$, i.e., $L_{R^{(3)}\delta K}$, has the second
order derivative couplings of scalar field to gravity which do
not appear in Horndeski theory or its extensions but does not
bring the Ostrogradski ghost. We discuss its implication on
scalar-tensor theory. We also show the interesting applications of
$L_{R^{(3)}\delta K}$ in the nonsingular cosmologies and the
inflation scenario.

\section{Higher order derivative coupling to gravity}

As was first found in \cite{Cai:2016thi} (see also
\cite{Creminelli:2016zwa}), the $R^{(3)}\delta g^{00}$ operator
plays a crucial role in solving the gradient instability
problem induced by $c_s^2<0$, (see also
\cite{deRham:2017aoj} for the unitarity problem), which 
suffered by the nonsingular cosmologies based on the Horndeski
theory
\cite{Libanov:2016kfc}\cite{Kobayashi:2016xpl}\cite{Kolevatov:2016ppi}\cite{Akama:2017jsa}.
In the Appendix, we point out that the $R^{(3)}\delta K$ operator
actually could play a role similar to that of $R^{(3)}\delta g^{00}$.
As will be shown, the covariant Lagrangian of $R^{(3)}\delta K$
contains the second order derivative of $\phi$ coupled to gravity,
such as $\sim{\phi^\mu\phi_{\mu\nu}\phi^\nu}R$, 
$\phi^\mu\phi^\nu(\Box\phi)R_{\mu\nu}$ and
${\phi^\mu\phi^\nu\phi_\rho\phi^{\rho\sigma}\phi_\sigma}R_{\mu\nu}$.
However, in Horndeski theory, such derivative couplings do
not appear, since they will bring the Ostrogradski ghost. Thus, it
is interesting to have a survey.

In this section, we will derive the covariant Lagrangian of
$R^{(3)}\delta K$ in unitary gauge. The induced metric on the
three-dimensional spacelike hypersurface ($\phi=const$) is
$h_{\mu\nu}=g_{\mu\nu}+n_{\mu}n_{\nu}$, where $n^\mu=-{1\over
\sqrt{-X}} \phi^\mu$ is the unit vector orthogonal to the
hypersurface and $n_{\mu}n^{\mu}=-1$, with $X=\phi_\mu\phi^\mu$
and $\phi^\mu=\nabla^\mu\phi$. The extrinsic curvature
$K_{\mu\nu}$ is defined as \be
K_{\mu\nu}=h_{\mu}^{\sigma}\nabla_{\sigma}n_{\nu}\,. \ee

Since $\delta K=K-3H$, it is straightforward to get \ba \delta K =
-{1\over \sqrt{-X}}\lf(\Box
\phi-{\phi^\mu\phi_{\mu\nu}\phi^\nu\over X}\rt)-3H\,, \ea with
$\phi_{\mu\nu}=\nabla_\nu\nabla_\mu\phi$. In unitary gauge
$\phi=\phi(t)$, we have $H=H(t(\phi))$. Using the Gauss-Codazzi
relation, we have \ba \label{RR3-cova} R^{(3)}&=&
R-{\phi_{\mu\nu}\phi^{\mu\nu}-(\Box \phi)^2\over X}
+{2\phi^\mu\phi_{\mu\nu}\phi^{\nu\sigma}\phi_\sigma \over
X^2}-{2\phi^\mu \phi_{\mu\nu}\phi^\nu \Box \phi\over X^2}
\nn\\&\,& -{2R_{\mu\nu}\phi^\mu\phi^\nu\over X}\,. \ea
Note that in the second line of Eq. (\ref{RR3-cova}), we also have
$R_{\mu\nu}\phi^\mu\phi^\nu=\phi_{\nu~\mu}^{~\mu~}
\phi^\nu-\phi^\nu_{~\nu\mu}\phi^\mu$ with
$\phi^\nu_{~\nu\mu}=\nabla_\mu\nabla_\nu \nabla^\nu\phi$, as given
in Ref. \cite{Cai:2017dyi}.

We define $S_{R^{(3)}\delta K}=\int d^4 x
\sqrt{-g}L_{R^{(3)}\delta K}$, with \ba \label{R3K}
L_{R^{(3)}\delta K}&=& \bar{f}_5\cdot \lf(R^{(3)}\delta K\rt)
\nn\\
&=& -{\bar{f}_5\over \sqrt{-X}}\lf[(\Box\phi) -
{\phi^\mu\phi_{\mu\nu}\phi^\nu\over
X}\rt]R\nn\\&\,&+{2\bar{f}_5\over
\sqrt{(-X)^3}}\lf[-\phi^\mu\phi^\nu(\Box\phi)
+{\phi^\mu\phi^\nu\phi_\rho\phi^{\rho\sigma}\phi_\sigma\over
X}\rt]R_{\mu\nu}\nn\\&\,& +{\bar{f}_5\over \sqrt{(-X)^{3}}}
\lf[(\Box\phi)^3-(\Box \phi)\phi_{\mu\nu}\phi^{\mu\nu}-
{(\Box\phi)^2\phi_\mu\phi^{\mu\nu}\phi_\nu-\phi_{\mu\nu}\phi^{\mu\nu}\phi_\rho\phi^{\rho\sigma}\phi_\sigma
\over X}\rt]\nn\\&\,& +{2\bar{f}_5\over \sqrt{(-X)^{5}}}\Big[
(\Box\phi)^2\phi_\mu\phi^{\mu\nu}\phi_\nu-(\Box\phi)
\phi_\mu\phi^{\mu\nu}\phi_{\nu\rho}\phi^\rho
\nn\\&\,&\quad\quad\quad\quad\quad\quad
-{(\Box\phi)(\phi_{\mu}\phi^{\mu\nu}\phi_\nu)^2-\phi_\mu\phi^{\mu\nu}\phi_{\nu\rho}\phi^{\rho}\phi_\sigma\phi^{\sigma\lambda}\phi_{\lambda}
\over X} \Big]-{\bar f}_4 R^{(3)}\,,
\ea where the leading contribution of $R^{(3)}\delta K$ is the
perturbation at quadratic order, so that $\bar{f}_5$ could be a
function of $\phi$, $X$ (and even $\Box \phi$ and
$\phi_\mu\phi^{\mu\nu}\phi_\nu$), and $\bar{f}_4=3\bar{f}_5
H(t(\phi))$. When $\bar{f}_4=0$ is set, $L_{R^{(3)}\delta K}$
reduces to $\sim R^{(3)}K$.


Recalling that in Horndeski theory, $L_5^H$ contains the 
coupling of the second order derivative of $\phi$ to 
gravity, i.e.,  $f(\phi,X)G_{\mu\nu}\phi^{\mu\nu}$ (or
$R_{\mu\nu}\phi^{\mu\nu}$).  Here, we require that $f$ is also $X$ dependent,
otherwise $G_{\mu\nu}\phi^{\mu\nu}$ will be
equivalent to $G_{\mu\nu}\phi^{\mu}\phi^{\nu}$, the cosmological
applications of which have been studied, see,
e.g., \cite{Feng:2013pba}\cite{Sadjadi:2013psa}\cite{Yang:2015pga}\cite{Harko:2016xip}.
While in $L_{R^{(3)}\delta K}$, the couplings \ba
(\Box\phi)R ,\quad & {\phi^\mu\phi_{\mu\nu}\phi^\nu}R,& \quad
 \phi^\mu\phi^\nu(\Box\phi)R_{\mu\nu}, \label{3}\\
&{\phi^\mu\phi^\nu\phi_\rho\phi^{\rho\sigma}\phi_\sigma}R_{\mu\nu}&
\label{4}\ea appear,  which are independent with
$R_{\mu\nu}\phi^{\mu\nu}$. In addition, such couplings to 
gravity also include
$\phi^{\mu\rho}\phi_\rho\phi^{\nu}R_{\mu\nu}$,
$\phi^{\mu}\phi^{\nu\rho}\phi^{\sigma}R_{\mu\nu\rho\sigma}$, which
are not independent and could be obtained by the combinations of
$R_{\mu\nu}\phi^{\mu\nu}$ and (\ref{3}),
as pointed out in Ref. \cite{BenAchour:2016fzp}.

In Horndeski theory, the cubic order of $\Box\phi$ in $L_5^H$ will
induce the higher derivatives in the metric and field equations,
which are actually set off by $G_{\mu\nu}\phi^{\mu\nu}$
\cite{Deffayet:2009mn}. This makes it be free from the Ostrogradski
ghost. In DHOST theory
\cite{Langlois:2015cwa}\cite{Langlois:2015skt}, all possible terms of cubic
order of the second order derivative of $\phi$ appear, which result in  higher order
equations of motion, but there is still no Ostrogradski
ghost due to the degeneracy.

Though the DHOST theory extends the Horndeski theory, the coupling
of the second order derivative of $\phi$ to gravity is still only
$G_{\mu\nu}\phi^{\mu\nu}$, since the derivative couplings
(\ref{3}) and (\ref{4}) will bring the Ostrogradski ghost (higher
derivatives in the equations of motion). However, in
$L_{R^{(3)}\delta K}$, the Ostrogradski ghost could be dispelled
by the combination of $(\Box\phi)^3$, $(\Box
\phi)\phi_{\mu\nu}\phi^{\mu\nu}$, etc., and $R^{(3)}$, see
(\ref{R3K}).

In principle, we could merge the Horndeski (even DHOST) theory and
$L_{R^{(3)}\delta K}$ into a (second order) derivative
coupling theory with all independent couplings
($R_{\mu\nu}\phi^{\mu\nu}$, (\ref{3}) and (\ref{4})) of the
second order derivative of $\phi$ to gravity. In such a theory,
the background equations of motion could be set only by the
Horndeski (DHOST) theory, since $L_{R^{(3)}\delta K}$ only
contributes $(\partial \zeta)^2,\,(\partial^2\zeta)^2$ at
leading order.

The quadratic coupling of the second order derivative of
$\phi$ to $R$, such as $(\Box\phi)^2R$, might be obtained in
$L\sim KR^{(3)}\delta K$ or equivalently ${\bar
f}_5(\Box\phi,\phi_\mu\phi^{\mu\nu}\phi_\nu)R^{(3)}\delta K$,
where all coefficients must be fixed as (\ref{R3K}). 

In mimetic gravity
\cite{Chamseddine:2014vna}\cite{Chamseddine:2016uef} (see e.g., \cite{Sebastiani:2016ras} for a review), since
$\delta g^{00}=0$, instead of $R^{(3)}\delta g^{00}$, the operator
$R^{(3)}\delta K$ might be significant for curing the
instabilities pointed out in
\cite{Ijjas:2016pad}\cite{Firouzjahi:2017txv}\cite{Hirano:2017zox}.
Here, since the mimetic constraint
$g^{\mu\nu}\phi_\mu\phi_\nu+1=0$ suggests $X=-1$, we have \be
R^{(3)}= {2 }\phi^\mu\phi^\nu R_{\mu\nu}+R -
\phi_{\mu\nu}\phi^{\mu\nu}-(\Box \phi)^2 \,, \ee the covariant
$L_{R^{(3)}\delta K}$ will be simpler.

It should be mentioned that at quadratic order $L_{R^{(3)}\delta
K}$ also contributes $(\partial^2\zeta)^2\sim
k^4\zeta^2$, which is harmful or harmless, depending on the
coefficient. However, $(\partial^2\zeta)^2$ could be
removed by using $(R^{(3)})^2$ (if required), since
$(R^{(3)})^2\sim (\partial^2\zeta)^2$ at leading order. We define
$S_{\lf(R^{(3)}\rt)^2}= \int d^4x\sqrt{-g} L_{\lf(R^{(3)}\rt)^2}$
and  $L_{\lf(R^{(3)}\rt)^2}=f_6\cdot\lf(R^{(3)}\rt)^2$, with \ba
\lf(R^{(3)}\rt)^2&=& R^2-{4\phi^\mu\phi^\nu R_{\mu\nu}R\over
X}+{4(\phi^\mu\phi^\nu R_{\mu\nu})^2\over X^2} \nn\\&\,&
+2R\lf[{(\Box\phi)^2-\phi_{\mu\nu}\phi^{\mu\nu}\over X}+{2 \phi^\mu\phi_{\mu\rho}\phi^{\rho\nu}\phi_\nu\over X^2}-{2(\Box\phi)\phi^\mu\phi_{\mu\nu}\phi^\nu\over X^2} \rt]
\nn\\&\,&
+4R_{\mu\nu}\phi^\mu\phi^\nu\lf[{\phi_{\rho\sigma}\phi^{\rho\sigma}\over X^2}
-{(\Box\phi)^2\over X^2}
-{2\phi_{\alpha}\phi^{\alpha\beta}\phi_{\beta\sigma}\phi^\sigma \over X^3}
+{2\Box\phi \phi_{\alpha}\phi^{\alpha\beta}\phi_{\beta} \over X^3}\rt]
\nn\\&\,&
+{(\phi_{\mu\nu}\phi^{\mu\nu})^2\over X^2}
-{4\phi_{\mu\nu}\phi^{\mu\nu} \phi_\alpha\phi^{\alpha\beta}\phi_{\beta\sigma}\phi^\sigma\over X^3}
+{4(\phi_\mu\phi^{\mu\nu}\phi_{\nu\rho}\phi^\rho)^2\over X^4}
\nn\\&\,&
+{4(\Box\phi)\phi_{\mu\nu}\phi^{\mu\nu}\phi_{\alpha}\phi^{\alpha\beta}\phi_{\beta}\over X^3}
-{8(\Box\phi)\phi_{\alpha}\phi^{\alpha\beta}\phi_{\beta} \phi_\mu\phi^{\mu\nu}\phi_{\nu\rho}\phi^\rho\over X^4}
\nn\\&\,&
+{4(\Box\phi)^2(\phi_\mu\phi^{\mu\nu}\phi_\nu)^2\over X^4}
-{2(\Box\phi)^2\phi_{\mu\nu}\phi^{\mu\nu}\over X^2 }
+{4(\Box\phi)^2\phi_\mu\phi^{\mu\nu}\phi_{\nu\rho}\phi^\rho\over X^3}
\nn\\&\,&
-{4(\Box\phi)^3\phi_\mu\phi^{\mu\nu}\phi_{\nu}\over X^3}
+{(\Box\phi)^4 \over X^2}\,, \ea where both $R^2$-order and the
coupling of $(\Box\phi)^2$ to $R$ actually appear and $f_6$ is a
function of $\phi$ and $X$ (and even $\Box \phi$ and
$\phi_\mu\phi^{\mu\nu}\phi_\nu$). In addition,
$L_{\lf(R^{(3)}\rt)^2}$ itself also has an interesting application in
nonsingular cosmologies
\cite{Cai:2016thi}\cite{Misonoh:2016btv}.

\section{Cosmological applications}

\subsection{Stable model for ekpyrotic scenario}

We consider the ekpyrotic scenario
\cite{Khoury:2001wf}\cite{Lehners:2008vx}. How to build a fully
stable bounce model is a significant issue. We proposed such a
model with $L_{R^{(3)}\delta g^{00}}$ in Ref. \cite{Cai:2017dyi}.
In Ref. \cite{Kolevatov:2017voe}, Kolevatov {\it et al.} also proposed a
different model by applying the ``inverse method" adopted in
\cite{Libanov:2016kfc}\cite{Ijjas:2016tpn}. However, with the
covariant $L_{R^{(3)}\delta g^{00}}$, the design is actually
simpler \cite{Cai:2017dyi}. Here, with $L_{R^{(3)}\delta K}$, the method is similar
(though slightly complicated).

We begin with the ekpyrotic Lagrangian \ba \label{action1} {\cal
L}_{ekpy}\sim & & \underbrace{{M_p^2\over2}R -X/2+{ V_0\over 2}
e^{\phi/{\cal M}_1
    }\left[1-\tanh( {\phi\over {\cal M}_2})\right]}\\ && \quad\quad   \text{\it Contraction\,\, and\,\, expansion}  \nn\\
    & &   \quad + \,\underbrace{{\tilde P}(\phi,X)}\quad (\text{around}\, \phi=0)\quad +  \quad \underbrace{L_{R^{(3)}\delta K} \quad or \quad
    L_{R^{(3)}\delta g^{00}} } \nn\\ &&
    \text{\it Bounce \,(NEC\, violation)} \quad  \quad \quad \quad\quad\quad \text{\it Removing} \,\, c_s^2<0\nn\\
 && \quad \text{\it Removing \, ghost} \nn\ea
with constant ${\cal M}_1, {\cal M}_2, V_0$. ${\tilde P}_X>1/2$
must be satisfied around $\phi\simeq 0$, so that ${\dot H}>0$. In
\cite{Cai:2017dyi}, see also
\cite{Koehn:2015vvy}\cite{Koehn:2013upa}, we adopted \be {\tilde
P}(\phi,X)={k_0\over (1+\kappa_1\phi^2)^2} {X/2}+ {q_0\over
(1+\kappa_2\phi^2)^2 }X^2\,\label{P} \ee with the constants $k_0,
\kappa_1$ (switching the sign before $X/2$ in (\ref{action1})
around $\phi\simeq 0$), and $q_0, \kappa_2$ (making $X^2$ appear
around $\phi\simeq 0$). A full ekpyrotic Lagrangian
(\ref{action1}) also should involve a mechanism (a coupling
$e^{-{\lambda\over M_p}\phi}\partial^\mu \chi\partial_{\mu}\chi$
\cite{Li:2013hga}\cite{Fertig:2013kwa}\cite{Ijjas:2014fja})
responsible for the scale invariant primordial perturbation.

The quadratic action of scalar perturbation for (\ref{action1}) is
\be S_{\zeta}^{(2)}=\int a^3 Q_s\lf(\dot{\zeta}^2-c_s^2
{(\partial\zeta)^2\over a^2} \rt)d^4x \,,\label{scalar-action} \ee
where \be Q_s={2{\dot \phi}^4{\tilde P}_{XX}-M_p^2{\dot H}\over
H^2},\quad c_s^2Q_s={{\dot c}_3\over a} -M_p^2, \label{cs2} \ee
and $c_3={aM_p^2\over H}(1-{2 \bar{f}_5 Q_s H\over M_p^4})$; see the
Appendix (or Ref. \cite{Cai:2016thi}) for details. In the Appendix, we
have $M_2^4(t)=\dot{\phi}^4\tilde{P}_{XX}$, ${\bar m}_5/2={\bar f}_5$ and
${\bar \lambda}/2={f}_6$. The quadratic action of tensor perturbation is
unaffected by $L_{R^{(3)}\delta K}$ and is still that in GR.

Here, we require $2X^2{\tilde P}_{XX}>M_p^2{\dot H}$, so that
$Q_s>0$ can be obtained.  If
$\bar{f}_5=0$, around the bounce point $H\simeq 0$,  we will have $c_s^2\sim -{\dot H}<0$. However, since
${\bar f}_5\neq 0$ and satisfies \be {2 \bar{f}_5 Q_s H\over
M_p^4}=1-{H\over aM_p^2}\int a\lf(Q_s c_s^2 +M_p^2\rt)dt,
\label{barf5}\ee we always could set $c_s^2\sim {\cal O}(1)$ with
suitable ${\bar f}_5$. It should be mentioned that when $H\sim 0$,
${\bar f}_5\sim {1\over HQ_s}\sim H$ crosses 0.

In (\ref{scalar-action}), $(\partial^2\zeta)^2$ has been canceled
by adding $L_{\lf(R^{(3)}\rt)^2}$ to ${\cal
L}_{ekpy}$ for simplicity, which requires \be 4f_6={{\bar
f}_5\over H}-\lf({3}+{Q_s\over M_p^2}\rt){{\bar f}_5^2\over
M_p^2}. \label{f6}\ee Thus, a fully stable nonsingular bounce
($Q_s>0$ and $c_s^2=1$) can be designed by using (\ref{action1})
with ${\bar f}_5$ given by (\ref{barf5}), and $f_4=3{\bar f}_5 H$,
and $f_6$ given by (\ref{f6}). With (\ref{P}), the calculation is
similar to that in Ref. \cite{Cai:2017dyi}.

\subsection{Slow-roll inflation with modified $c_s^2$ }

We consider the inflation scenario. Here, the covariant
$L_{R^{(3)}\delta K}$ and also $L_{R^{(3)}\delta g^{00}}$ only
affect the sound speed $c_s$ of scalar perturbation, but the
background and the tensor perturbation are unaffected. The effect
of modified $c_s^2$ may be encoded in the power spectrum of
primordial scalar perturbation, which might be observable.

The Lagrangian is \ba \label{act02} {\cal L}\sim {M_p^2\over
2}R+L_{inf}+L_{R^{(3)}\delta K}+L_{(R^{(3)})^2}\, , \ea where
$L_{inf}=-\phi_\mu\phi^\mu/2-V(\phi)$ is responsible for the
inflation. We set the slow-roll parameter $\epsilon=-{\dot{H}/
H^2}=const>0$ for simplicity. The quadratic action of scalar
perturbation is given in (\ref{SS}) of the Appendix with $M_2={\tilde
m}_4=0$. We have $Q_s=\epsilon M_p^2$ and  \be c_s^2=
1-{\bar{m}_5H\over M_p^2}-{\dot{\bar{m}}_5\over
M_p^2}\,,\label{css}\ee \be c_4\simeq {3{\bar m}_5^2\over
M_p^2}-{2{\bar m}_5\over H}+8{\bar \lambda}. \label{c4}\ee Here,
$L_{R^{(3)}\delta K}$ modifies $c_s^2$. We require $c_4=0$, which
suggests that $\bar{\lambda}$ in (\ref{c4}) is determined by
$\bar{m}_5$ and $H$.

The equation of motion for $\zeta$ is \be u''+\lf({c}_{s}^2
k^2-{z_s''\over z_s} \rt)u=0 \label{MSEQ} \ee with the
definition $u=z_s\zeta$ and $z_s=\sqrt{2a^2 \epsilon M_p^2}$, and
the superscript $^\prime$ is the derivative with respect to
$\tau=\int dt/a$. The initial state of the perturbation mode is $u=
{1\over\sqrt{2c_sk}}e^{-ic_sk\tau}$.
The power spectrum of $\zeta$ is \be P_{\cal R}={k^3\over
2\pi^2}\lf|{u\over z_s} \rt|^2 \,.\ee We have $P_{\cal
R}^{inf}=\frac{H^2_{inf}}{8 \pi ^2 M_p^2 \epsilon }\lf({k\over aH}
\rt)^{-2\epsilon}$ for slow-roll inflation ($c_s^2=1$). Here, if
$c_s^2=const<1$ is required, ${\dot {\bar m}}_5=0$ in (\ref{css})
should be satisfied. This will result in
$c_s^2=1-{\bar{m}_5H_{inf}\over M_p^2}\simeq 1$, since $H_{inf}\ll
M_p$ while ${\bar m}_5\lesssim M_p$. Thus, the case with $c_s^2\neq
const$ might be interesting.

\begin{figure}[htbp]
    \subfigure[~~$c_s^2$]{\includegraphics[width=.45\textwidth]{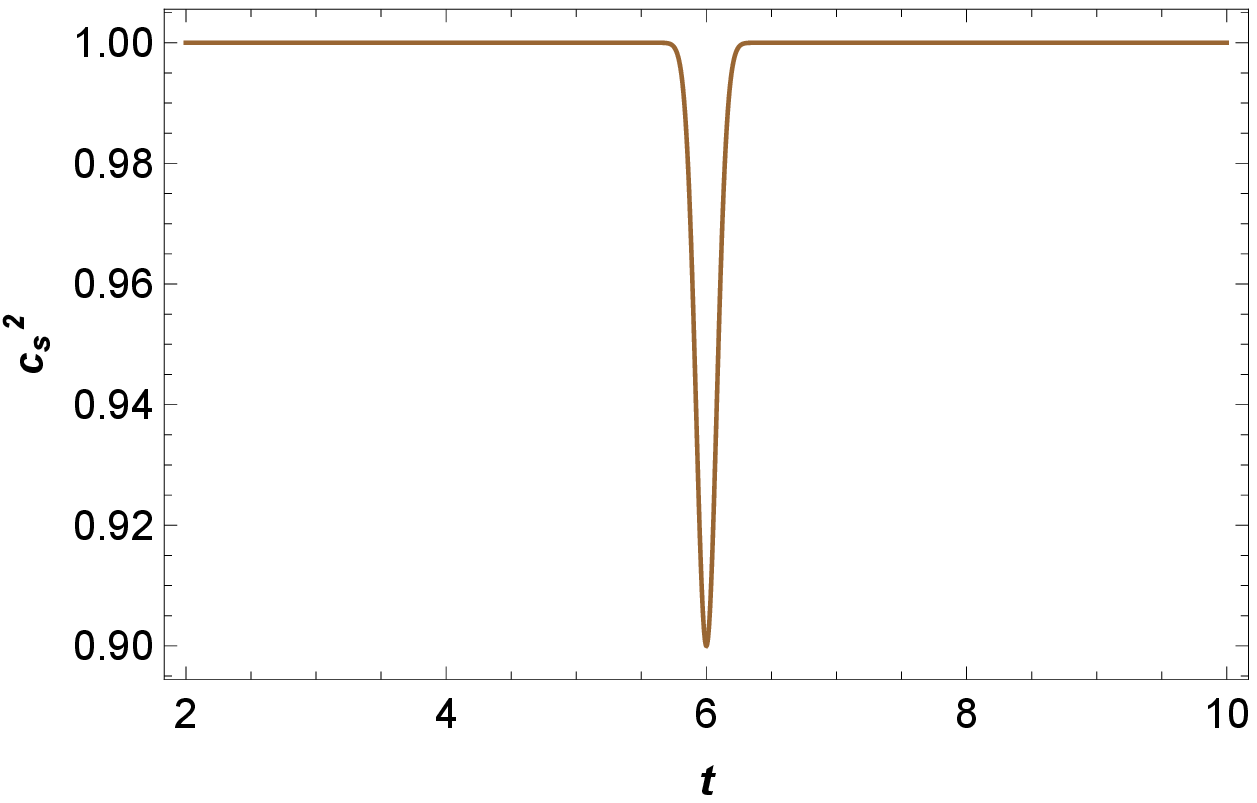} }
    \subfigure[~~$\bar{m}_5$ and $\dot{\bar{m}}_5$]{\includegraphics[width=.46\textwidth]{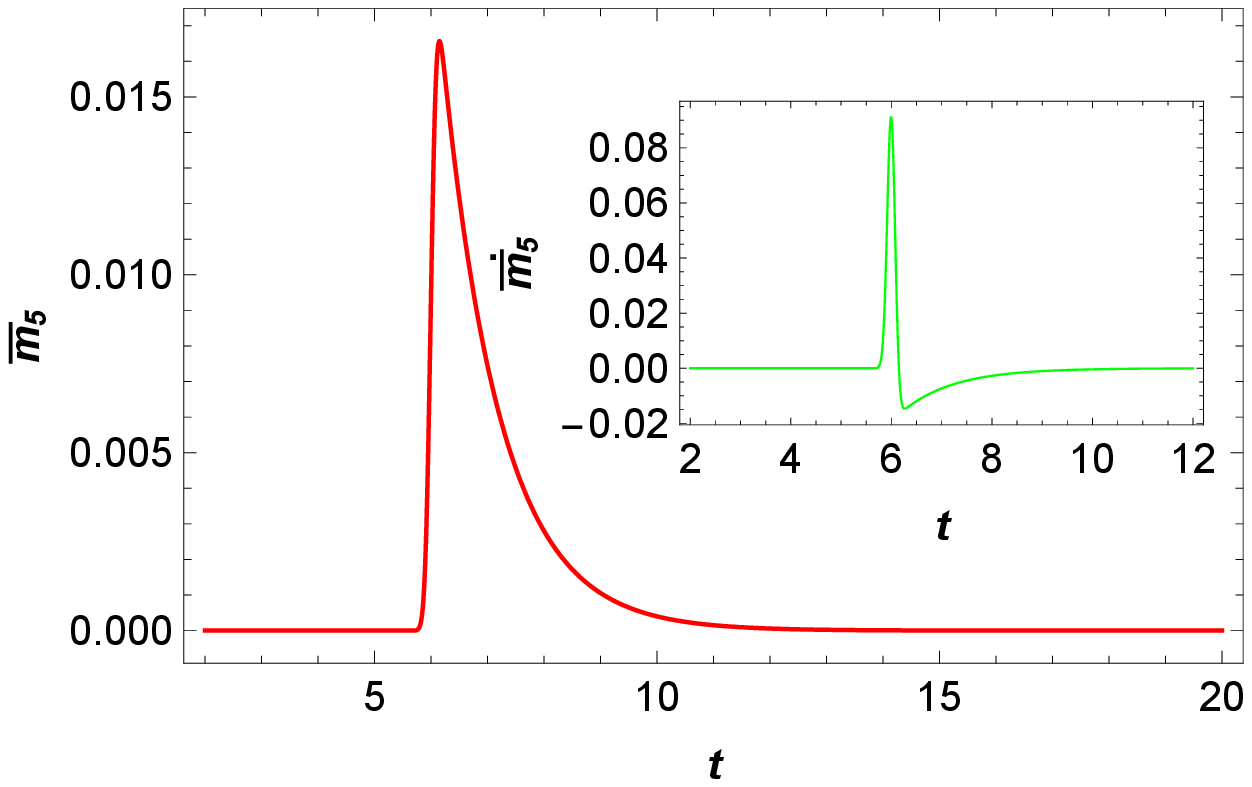} }
    \subfigure[~~$\bar{\lambda}$]{\includegraphics[width=.45\textwidth]{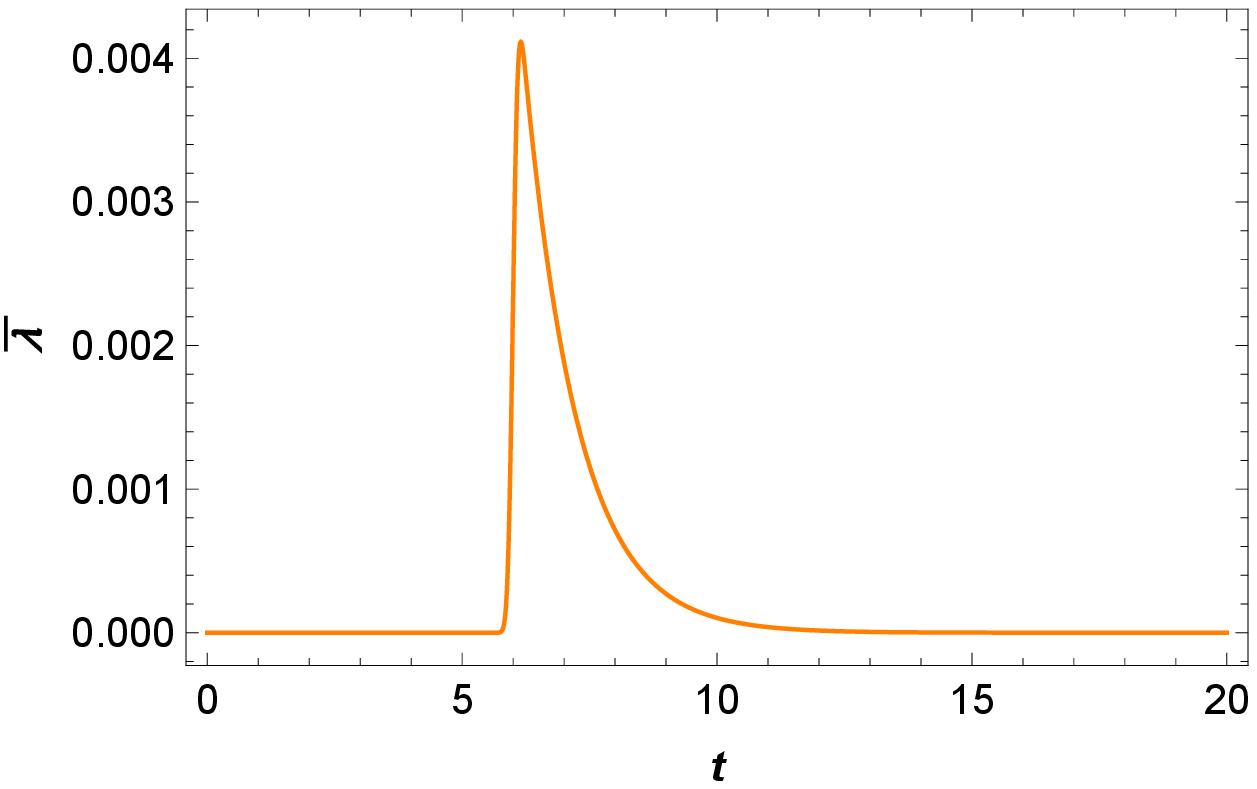} }
    \subfigure[~~$P_{\cal R}/P_{\cal R}^{inf}$]{\includegraphics[width=.45\textwidth]{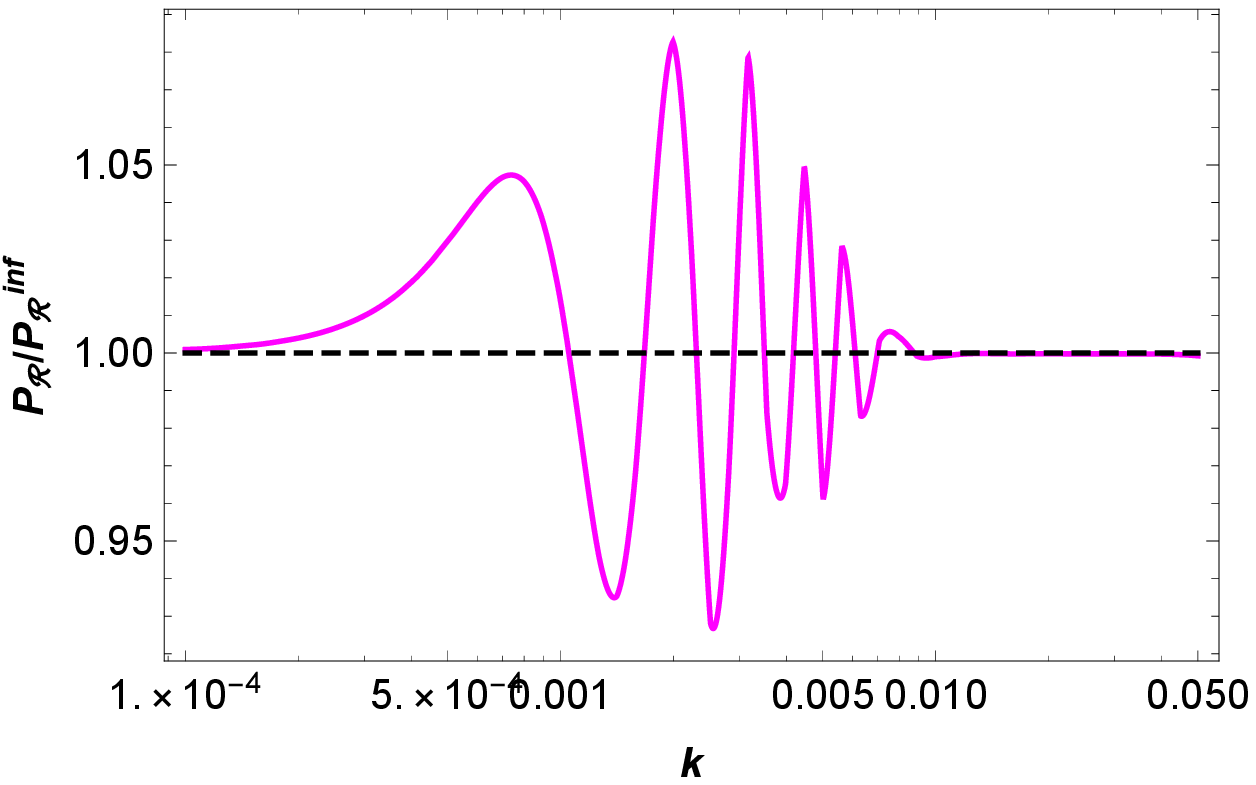} }
    \caption{
    The background is the slow-roll
    inflation with $\epsilon=0.003$. We set $c_s^2=1 - {\cal A}_* e^{-{\cal B}_*(t - t_*)^2}$ with ${\cal A}_*=0.1$,
    ${\cal B}_*=80$ and $t_*=6$.  } \label{fig02}
\end{figure}

For an example, we consider a model in which $c_s^2$
acquires a dip (Fig.\ref{fig02}(a)). We numerically show the
corresponding evolutions of $\bar{m}_5$ and $\bar{\lambda}$ in
Figs. \ref{fig02}(b) and \ref{fig02}(c), according to (\ref{css}) and
(\ref{c4}), which could be rewritten as $\bar{m}_5(\phi)$ and
$\bar{\lambda}(\phi)$ since $\phi=\phi(t)$. We plot $P_{\zeta}$ in
Fig.\ref{fig02} (d) by solving Eq.(\ref{MSEQ}), see
\cite{Cai:2015yza}\cite{Cai:2015dta}\cite{Cai:2015ipa} for a similar
method. We see that the effect of $L_{R^{(3)}\delta K}$ on $c_s^2$
could be encoded in the power spectrum of scalar perturbation.

The phenomenological effect of $L_{R^{(3)}\delta K}$ is very similar to that of $L_{R^{(3)}\delta
		g^{00}}$ at
	quadratic order, if the contribution of $L_{R^{(3)}\delta K}$ to
		term $\sim k^4\zeta^2$ in the quadratic action is totally canceled by
		$L_{\lf(R^{(3)}\rt)^2}$, i.e., $c_4=0$, which requires
	$\bar{\lambda}=\bar{\lambda}_0$ with $\bar{\lambda}_0\simeq
	{3{\bar m}_5^2\over
		8M_p^2}-{{\bar m}_5\over 4H}$. However, when the condition $c_4=0$ is violated, Eq. (\ref{MSEQ}) should be modified to
	$u''+\lf({c}_{s,eff}^2
	k^2-{z_s''/ z_s} \rt)u=0$
	where $c_{s,eff}^2=c_s^2-2c_4k^2/z_s^2$ (for simplicity, we will focus on the cases in which $c_4=0$ initially so that the initial state of the perturbation mode is still $u=
	{1\over\sqrt{2c_sk}}e^{-ic_sk\tau}$).

\begin{figure}[htbp]
	\subfigure[~~]{\includegraphics[width=.46\textwidth]{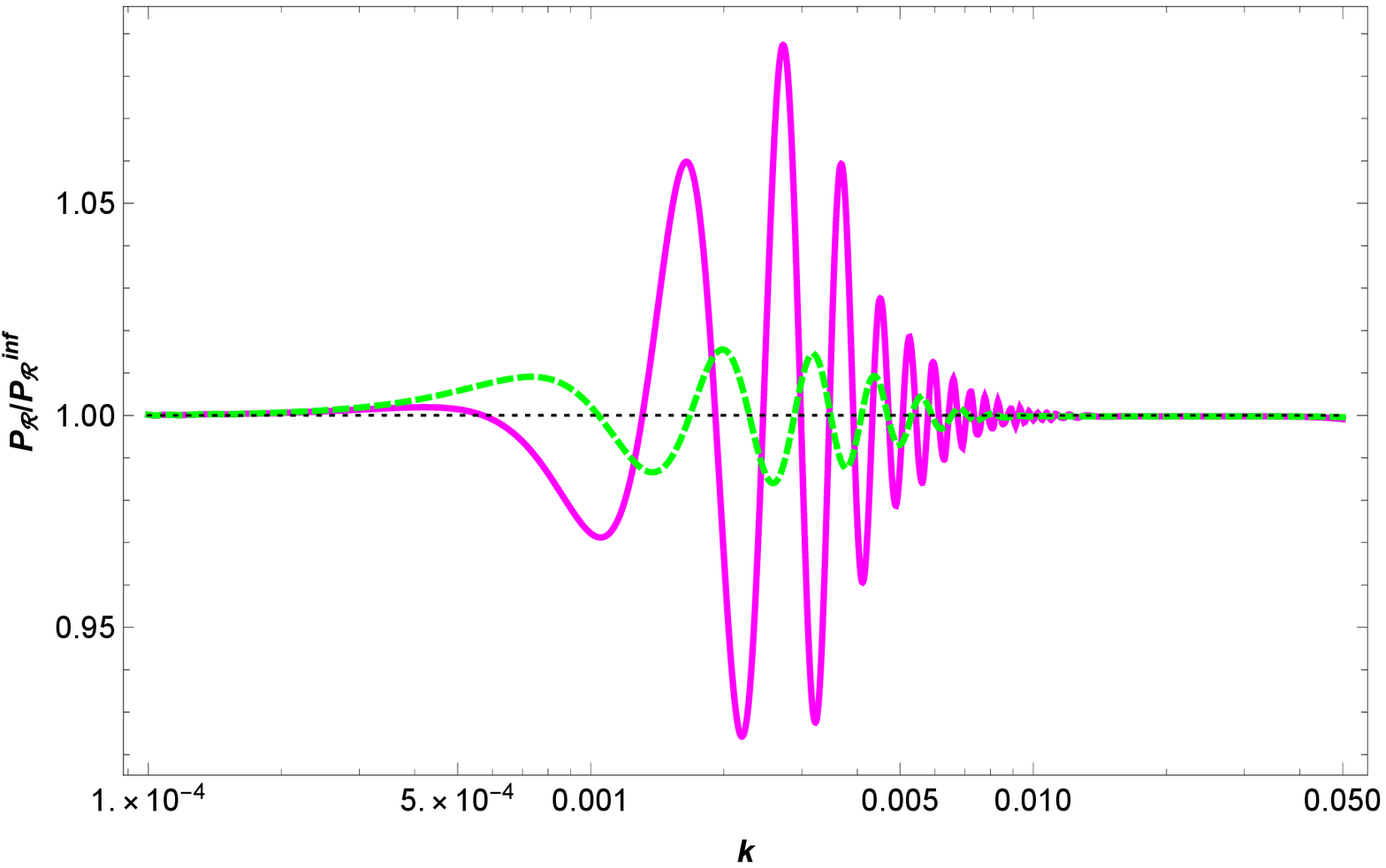} }
	\subfigure[~~]{\includegraphics[width=.46\textwidth]{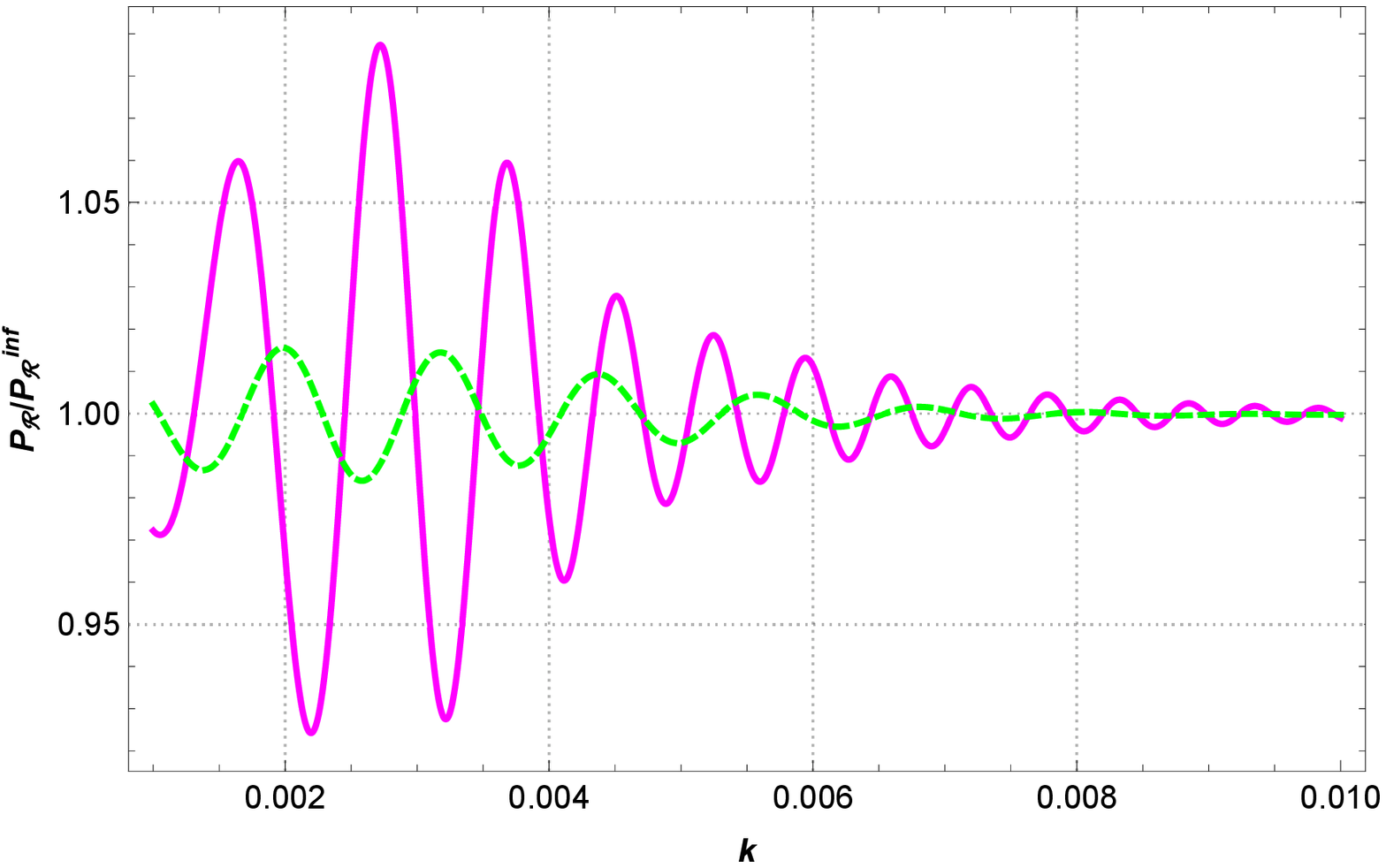} }
	\caption{
		The background is the slow-roll
		inflation with $\epsilon=0.003$. We set $c_s^2=1 - {\cal A}_* e^{-{\cal B}_*(t - t_*)^2}$ with ${\cal A}_*=0.02$,
		${\cal B}_*=80$ and $t_*=6$ for both (a) and (b), while we set $\bar{\lambda}=\bar{\lambda}_0$ (i.e., $c_4=0$) for the green dashed curves and $\bar{\lambda}=0.997\bar{\lambda}_0$ (i.e., $c_4\neq 0$) for the magenta solid curves. } \label{fig03}
\end{figure}

Phenomenologically, we could 
		distinguish the operator $L_{R^{(3)}\delta K}$ from
		$L_{R^{(3)}\delta g^{00}}$. First, when $c_4\neq0$, the frequency of the
		oscillations in the power spectrum will increase with $k$,
		while the frequency of the oscillations is nearly constant for $c_4=0$,
		see Fig. \ref{fig03}(b). Second, when $c_4\neq0$ (even when $\bar{\lambda}$ slightly
		deviates from $\bar{\lambda}_0$), $c_{s,eff}^2$ may induce a larger
		amplitude of oscillations than that of $c_s^2$ in the power
		spectrum, as numerically shown in Fig. \ref{fig03}, unless $c_s^2$ has more
		drastic (or fine-tuned) variation.

The effect of varying $c_s^2$ on scalar perturbations has been
also studied in Refs.
\cite{Nakashima:2010sa}\cite{Park:2012rh}\cite{Bartolo:2013exa}\cite{Achucarro:2014msa}\cite{Saito:2013aqa}\cite{Mizuno:2014jja},
but based on $P(\phi,X)$ (or equivalent EFT).

\section{Discussion}

Recently, it has been found in
\cite{Cai:2016thi}\cite{Creminelli:2016zwa} that the operators
with three-dimensional Ricci scalar $R^{(3)}$ in EFT, especially
$R^{(3)}\delta g^{00}$, are significant for solving the problem of
$c_s^2<0$, which is suffered by the nonsingular cosmologies. Here, we find
that the $R^{(3)}\delta K$ operator actually could play a role similar to that of
$R^{(3)}\delta g^{00}$.

We derived the covariant Lagrangian of $R^{(3)}\delta K$. The
covariant $L_{R^{(3)}\delta K}$ has the second order derivative
coupling of the field $\phi$ to  gravity, such as (\ref{3})
and (\ref{4}) (which do not appear in Horndeski and DHOST theory),
but does not bring the Ostrogradski ghost. This suggests that the
Horndeski (or even  DHOST) theory and $L_{R^{(3)}\delta K}$
might be merged into a second order derivative coupling
theory with all possible independent couplings, i.e.,
$G_{\mu\nu}\phi^{\mu\nu}$ (or $R_{\mu\nu}\phi^{\mu\nu}$),
(\ref{3}) and (\ref{4}), of the second order derivative of $\phi$
to gravity. Here, how (\ref{3}) and (\ref{4}) consistently appear
in such a theory is just what is told by the covariant description of
the $R^{(3)}\delta K$ operator.

With $L_{R^{(3)}\delta K}$, we built a fully stable cosmological
model for the ekpyrotic scenario, by applying similar method used in
Ref. \cite{Cai:2017dyi}.
Our work indicates that with the covariant $L_{R^{(3)}\delta
g^{00}}$ (proposed in \cite{Cai:2017dyi}) or $L_{R^{(3)}\delta
K}$, the stable nonsingular bounce scenario could be concisely
designed. Here, our study is motivated straightly by the EFT
operators, e.g., \cite{Cai:2016thi}. However, other studies based
on modified gravity will also be interesting
\cite{Banerjee:2016hom,Biswas:2011ar,Vasilic:2017myy,Li:2016xjb,Odintsov:2015zza,Hendi:2016tiy,Giovannini:2016jkf},
especially their stabilities.

We also studied the possible effect of $L_{R^{(3)}\delta K}$ on
the primordial scalar perturbation in the inflation scenario, which
might be encoded in the TT spectrum of cosmic microwave background (CMB). We will come back to the
relevant issues elsewhere.

\textbf{Acknowledgments}

We thank Yunlong Zheng, Mingzhe Li and Xian Gao for helpful
discussions. YC would like to thank Youping Wan and Yi-Fu Cai for
discussions and hospitality during his visit at University of
Science and Technology of China. This work is supported by NSFC,
No. 11575188, 11690021, and also supported by the Strategic
Priority Research Program of CAS, No. XDA04075000, XDB23010100.


\appendix

\section{The EFT }

As pointed out in Refs. \cite{Ijjas:2016tpn}\cite{Cai:2017tku}, the
cubic Galileon only moves the period of $c_s^2<0$ to the outside
of the null energy condition (NEC) violating phase but {cannot} dispel it completely, see also the
earlier discussion \cite{Easson:2011zy} on this point.

In this Appendix, we briefly review the EFT for nonsingular
cosmologies, and show how the $R^{(3)}\delta g^{00}$ and
$R^{(3)}\delta K$ operators play crucial roles in solving the
problem of $c_s^2<0$.

With the ADM line element, we have 
\begin{equation}
g_{\mu\nu}=\left(
  \begin{array}{cc}
  N_kN^k-N^2 &  N_j\\
  N_i &  h_{ij}\\
  \end{array}
\right) \,,\qquad
g^{\mu\nu}=\left(
  \begin{array}{cc}
  -N^{-2} &  {N^j\over N^2}\\
  {N^i\over N^2} &  h^{ij}-{N^iN^j\over N^2}\\
  \end{array}
\right) \,,\qquad
\end{equation}
and $\sqrt{-g}=N\sqrt{h}$, where $N_i=h_{ij}N^j$.  The induced metric on
three-dimensional hypersurface is
$h_{\mu\nu}=g_{\mu\nu}+n_{\mu}n_{\nu}$, where $n_{\mu}=n_0
(dt/dx^{\mu})=(-N,0,0,0)$, $n^{\nu}=g^{\mu\nu}n_{\mu} =({1/ N},-{N^i/ N})$ is orthogonal to the spacelike hypersurface, and $n_{\mu}n^{\mu}=-1$. Thus,
\begin{equation}
h_{\mu\nu}=\left(
  \begin{array}{cc}
  N_kN^k &  N_j\\
  N_i &  h_{ij}\\
  \end{array}
\right) \,,\qquad
h^{\mu\nu}=\left(
  \begin{array}{cc}
  0 &  0\\
  0 &  h^{ij}\\
  \end{array}
\right) \,.\qquad
\end{equation}

The EFT action is \ba \label{eft_action} S&=&\int d^4x\sqrt{-g}\Big[
{M_p^2\over2} f(t)R-\Lambda(t)-c(t)g^{00}
\nn\\
&\,&+{M_2^4(t)\over2}(\delta g^{00})^2-{m_3^3(t)\over2}\delta
K\delta g^{00} -m_4^2(t)\lf( \delta K^2-\delta K_{\mu\nu}\delta
K^{\mu\nu} \rt) +{\tilde{m}_4^2(t)\over 2}R^{(3)}\delta g^{00}
\nn\\
&\,&-\bar{m}_4^2(t)\delta K^2+{\bar{m}_5(t)\over 2}R^{(3)}\delta K
+{\bar{\lambda}(t)\over2}(R^{(3)})^2+...
\nn\\
&\,& -{\tilde{\lambda}(t)\over
    M_p^2}\nabla_iR^{(3)}\nabla^iR^{(3)} +... \Big] \,,
\ea where $\delta g^{00}=g^{00}+1$, $R^{(3)}$ is the three-dimensional
Ricci scalar, $K_{\mu\nu}=h_{\mu}^{\sigma}\nabla_{\sigma}n_{\nu}$
is the extrinsic curvature and $\delta
K_{\mu\nu}=K_{\mu\nu}-h_{\mu\nu}H$. The  first row describes the
background, while the rest are for the perturbations. We always
could set $f=1$, which implies $c(t)=-M_p^2{\dot H}$ and
$c(t)+\Lambda(t)=3M_p^2H^2$. See, e.g., \cite{Cai:2016thi} for the
details.

Here, we only consider the coefficients set $(M_2, {\tilde m}_4,
{\bar m}_5, {\bar \lambda})$ and set other coefficients in
(\ref{eft_action}) equal to 0. Only with $(M_2, {\tilde m}_4,
{\bar m}_5, {\bar \lambda})\neq 0$, the quadratic action of scalar
perturbation $\zeta$ is (see, e.g., our \cite{Cai:2016thi}) \ba
\label{SS} S^{(2)}_\zeta=\int d^4x\,a^3Q_s\lf[
\dot{\zeta}^2-c_s^2{(\partial
    \zeta)^2\over a^2}+{c_4\over a^4Q_s}(\partial^2\zeta)^2
\rt]\,, \ea where \ba &\,&Q_s=\frac{2 M_2^4}{H^2}-\frac{\dot{H}
M_p^2}{H^2}\,,
\\ &\,&
c^2_sQ_s={\dot{c}_3\over a}-c_2 \,
\\
&\,&c_2=M_p^2\,,
\\
&\,&c_3=-\frac{2 a M_2^4 \bar{m}_5}{H^2 M_p^2}+\frac{a \dot{H}
    \bar{m}_5}{H^2}+\frac{a M_p^2}{H}+{ 2a{\tilde m}_4^2\over H}\,,
\\
&\,&c_4=\frac{2 M_2^4 \bar{m}_5^2}{H^2 M_p^4}-\frac{\dot{H}
    \bar{m}_5^2}{H^2 M_p^2}-\frac{2 \bar{m}_5}{H}+\frac{3
    \bar{m}_5^2}{M_p^2}
 -{4{\bar m}_5{\tilde m}_4^2\over HM_p^2}+8\bar{\lambda}\,. \ea

Only if $Q_s>0$ and $c_s^2>0$ is the nonsingular cosmological model healthy. In models with the operator $(\delta g^{00})^2$,
$Q_s>0$ can be obtained, since $(\delta g^{00})^2$ contributes
${\dot \zeta}^2$, while $c_s^2<0$ can be avoided since
$R^{(3)}\delta g^{00}$ or $R^{(3)}\delta K$ contributes
$(\partial\zeta)^2$.


\begin{thebibliography}{99}

\bibitem{Rubakov:2014jja}
  V.~A.~Rubakov,
  Phys.\ Usp.\  {\bf 57}, 128 (2014)
  [Usp.\ Fiz.\ Nauk {\bf 184}, 2, 137 (2014)]
  [arXiv:1401.4024 [hep-th]].

\bibitem{Joyce:2014kja}
  A.~Joyce, B.~Jain, J.~Khoury and M.~Trodden,
  Phys.\ Rept.\  {\bf 568}, 1 (2015)
  [arXiv:1407.0059 [astro-ph.CO]].

\bibitem{Nojiri:2017ncd}
  S.~Nojiri, S.~D.~Odintsov and V.~K.~Oikonomou,
  arXiv:1705.11098 [gr-qc].

\bibitem{Horndeski:1974wa}
G.~W.~Horndeski,
Int.\ J.\ Theor.\ Phys.\  {\bf 10}, 363 (1974).

\bibitem{Deffayet:2011gz}
C.~Deffayet, X.~Gao, D.~A.~Steer and G.~Zahariade,
Phys.\ Rev.\ D {\bf 84}, 064039 (2011)
[arXiv:1103.3260 [hep-th]].


\bibitem{Kobayashi:2011nu}
T.~Kobayashi, M.~Yamaguchi and J.~Yokoyama,
Prog.\ Theor.\ Phys.\  {\bf 126}, 511 (2011)
[arXiv:1105.5723 [hep-th]].

\bibitem{Zumalacarregui:2013pma}
M.~Zumalacárregui and J.~García-Bellido,
Phys.\ Rev.\ D {\bf 89}, 064046 (2014)
[arXiv:1308.4685 [gr-qc]].

\bibitem{Gleyzes:2014dya}
J.~Gleyzes, D.~Langlois, F.~Piazza and F.~Vernizzi,
Phys.\ Rev.\ Lett.\  {\bf 114}, 21, 211101 (2015)
[arXiv:1404.6495 [hep-th]].

\bibitem{Gleyzes:2014qga}
J.~Gleyzes, D.~Langlois, F.~Piazza and F.~Vernizzi,
JCAP {\bf 1502}, 018 (2015)
[arXiv:1408.1952 [astro-ph.CO]].

\bibitem{Langlois:2015cwa}
D.~Langlois and K.~Noui,
JCAP {\bf 1602}, 02, 034 (2016) 
[arXiv:1510.06930 [gr-qc]].

\bibitem{Langlois:2015skt}
D.~Langlois and K.~Noui,
JCAP {\bf 1607}, 07, 016 (2016)
[arXiv:1512.06820 [gr-qc]].

\bibitem{BenAchour:2016fzp}
J.~Ben Achour, M.~Crisostomi, K.~Koyama, D.~Langlois, K.~Noui and G.~Tasinato,
JHEP {\bf 1612}, 100 (2016)
[arXiv:1608.08135 [hep-th]].


\bibitem{Langlois:2017mxy}
D.~Langlois, M.~Mancarella, K.~Noui and F.~Vernizzi,
JCAP {\bf 1705}, 05, 033 (2017)
[arXiv:1703.03797 [hep-th]].

\bibitem{Cheung:2007st}
C.~Cheung, P.~Creminelli, A.~L.~Fitzpatrick, J.~Kaplan and L.~Senatore,
JHEP {\bf 0803}, 014 (2008)
[arXiv:0709.0293 [hep-th]].

\bibitem{Weinberg:2008hq}
S.~Weinberg,
Phys.\ Rev.\ D {\bf 77}, 123541 (2008)
[arXiv:0804.4291 [hep-th]].

\bibitem{Gubitosi:2012hu}
G.~Gubitosi, F.~Piazza and F.~Vernizzi,
JCAP {\bf 1302}, 032 (2013)
[arXiv:1210.0201 [hep-th]].

\bibitem{Gleyzes:2013ooa}
J.~Gleyzes, D.~Langlois, F.~Piazza and F.~Vernizzi,
JCAP {\bf 1308}, 025 (2013)
[arXiv:1304.4840 [hep-th]].

\bibitem{Piazza:2013coa}
F.~Piazza and F.~Vernizzi,
Class.\ Quant.\ Grav.\  {\bf 30}, 214007 (2013)
[arXiv:1307.4350 [hep-th]].

\bibitem{Naskar:2017ekm}
A.~Naskar, S.~Choudhury, A.~Banerjee and S.~Pal,
arXiv:1706.08051 [astro-ph.CO].

\bibitem{Cai:2016thi}
Y.~Cai, Y.~Wan, H.~G.~Li, T.~Qiu and Y.~S.~Piao,
JHEP {\bf 1701}, 090 (2017)
[arXiv:1610.03400 [gr-qc]].

\bibitem{Creminelli:2016zwa}
P.~Creminelli, D.~Pirtskhalava, L.~Santoni and E.~Trincherini,
JCAP {\bf 1611}, 11, 047 (2016)
[arXiv:1610.04207 [hep-th]].

\bibitem{Cai:2017tku}
Y.~Cai, H.~G.~Li, T.~Qiu and Y.~S.~Piao,
Eur.\ Phys.\ J.\ C {\bf 77}, 6, 369 (2017)
[arXiv:1701.04330 [gr-qc]].



\bibitem{Libanov:2016kfc}
M.~Libanov, S.~Mironov and V.~Rubakov,
JCAP {\bf 1608}, 08, 037 (2016)
[arXiv:1605.05992 [hep-th]].


\bibitem{Kobayashi:2016xpl}
T.~Kobayashi,
Phys.\ Rev.\ D {\bf 94}, 4, 043511 (2016)
[arXiv:1606.05831 [hep-th]].


\bibitem{Cai:2017dyi}
Y.~Cai and Y.~S.~Piao,
arXiv:1705.03401 [gr-qc].


\bibitem{Ijjas:2016vtq}
A.~Ijjas and P.~J.~Steinhardt,
Phys.\ Lett.\ B {\bf 764}, 289 (2017)
[arXiv:1609.01253 [gr-qc]].


\bibitem{Yoshida:2017swb}
D.~Yoshida, J.~Quintin, M.~Yamaguchi and R.~H.~Brandenberger,
arXiv:1704.04184 [hep-th].


\bibitem{Kolevatov:2017voe}
R.~Kolevatov, S.~Mironov, N.~Sukhov and V.~Volkova,
arXiv:1705.06626 [hep-th].


\bibitem{Zheng:2017qfs}
Y.~Zheng, L.~Shen, Y.~Mou and M.~Li,
arXiv:1704.06834 [gr-qc].


\bibitem{deRham:2017aoj}
C.~de Rham and S.~Melville,
Phys.\ Rev.\ D {\bf 95}, no. 12, 123523 (2017)
[arXiv:1703.00025 [hep-th]].


\bibitem{Kolevatov:2016ppi}
R.~Kolevatov and S.~Mironov,
Phys.\ Rev.\ D {\bf 94}, 12, 123516 (2016)
[arXiv:1607.04099 [hep-th]].



\bibitem{Akama:2017jsa}
S.~Akama and T.~Kobayashi,
Phys.\ Rev.\ D {\bf 95}, 6, 064011 (2017)
[arXiv:1701.02926 [hep-th]].




\bibitem{Feng:2013pba}
K.~Feng, T.~Qiu and Y.~S.~Piao,
Phys.\ Lett.\ B {\bf 729}, 99 (2014)
[arXiv:1307.7864 [hep-th]].

\bibitem{Sadjadi:2013psa}
H.~Mohseni Sadjadi and P.~Goodarzi,
Phys.\ Lett.\ B {\bf 732}, 278 (2014)
[arXiv:1309.2932 [astro-ph.CO]].

\bibitem{Yang:2015pga}
N.~Yang, Q.~Fei, Q.~Gao and Y.~Gong,
Class.\ Quant.\ Grav.\  {\bf 33}, 20, 205001 (2016)
[arXiv:1504.05839 [gr-qc]];
Y.~Zhu and Y.~Gong,
Int.\ J.\ Mod.\ Phys.\ D {\bf 26}, 02, 1750005 (2016)
[arXiv:1512.05555 [gr-qc]].


\bibitem{Harko:2016xip}
T.~Harko, F.~S.~N.~Lobo, E.~N.~Saridakis and M.~Tsoukalas,
Phys.\ Rev.\ D {\bf 95}, 4, 044019 (2017)
[arXiv:1609.01503 [gr-qc]];
J.~B.~Dent, S.~Dutta, E.~N.~Saridakis and J.~Q.~Xia,
JCAP {\bf 1311}, 058 (2013)
[arXiv:1309.4746 [astro-ph.CO]].


\bibitem{Deffayet:2009mn}
  C.~Deffayet, S.~Deser and G.~Esposito-Farese,
  Phys.\ Rev.\ D {\bf 80}, 064015 (2009)
  [arXiv:0906.1967 [gr-qc]].


\bibitem{Chamseddine:2014vna}
A.~H.~Chamseddine, V.~Mukhanov and A.~Vikman,
JCAP {\bf 1406}, 017 (2014)
[arXiv:1403.3961 [astro-ph.CO]].



\bibitem{Chamseddine:2016uef}
A.~H.~Chamseddine and V.~Mukhanov,
JCAP {\bf 1703}, 03, 009 (2017)
[arXiv:1612.05860 [gr-qc]].

\bibitem{Sebastiani:2016ras}
L.~Sebastiani, S.~Vagnozzi and R.~Myrzakulov,
Adv.\ High Energy Phys.\  {\bf 2017}, 3156915 (2017)
[arXiv:1612.08661 [gr-qc]].

\bibitem{Ijjas:2016pad}
A.~Ijjas, J.~Ripley and P.~J.~Steinhardt,
Phys.\ Lett.\ B {\bf 760}, 132 (2016)
[arXiv:1604.08586 [gr-qc]].


\bibitem{Firouzjahi:2017txv}
H.~Firouzjahi, M.~A.~Gorji and A.~Hosseini Mansoori,
arXiv:1703.02923 [hep-th].


\bibitem{Hirano:2017zox}
S.~Hirano, S.~Nishi and T.~Kobayashi,
arXiv:1704.06031 [gr-qc].





\bibitem{Misonoh:2016btv}
Y.~Misonoh, M.~Fukushima and S.~Miyashita,
Phys.\ Rev.\ D {\bf 95}, 4, 044044 (2017)
[arXiv:1612.09077 [gr-qc]].




\bibitem{Khoury:2001wf}
J.~Khoury, B.~A.~Ovrut, P.~J.~Steinhardt and N.~Turok,
Phys.\ Rev.\ D {\bf 64}, 123522 (2001)
[hep-th/0103239].

\bibitem{Lehners:2008vx}
J.~L.~Lehners,
Phys.\ Rept.\  {\bf 465}, 223 (2008)
[arXiv:0806.1245 [astro-ph]].





\bibitem{Ijjas:2016tpn}
A.~Ijjas and P.~J.~Steinhardt,
Phys.\ Rev.\ Lett.\  {\bf 117}, 12, 121304 (2016)
[arXiv:1606.08880 [gr-qc]].

\bibitem{Koehn:2015vvy}
M.~Koehn, J.~L.~Lehners and B.~Ovrut,
Phys.\ Rev.\ D {\bf 93}, 10, 103501 (2016)
[arXiv:1512.03807 [hep-th]].





\bibitem{Koehn:2013upa}
M.~Koehn, J.~L.~Lehners and B.~A.~Ovrut,
Phys.\ Rev.\ D {\bf 90}, 2, 025005 (2014)
[arXiv:1310.7577 [hep-th]].


\bibitem{Li:2013hga}
M.~Li,
Phys.\ Lett.\ B {\bf 724}, 192 (2013)
[arXiv:1306.0191 [hep-th]].

\bibitem{Fertig:2013kwa}
A.~Fertig, J.~L.~Lehners and E.~Mallwitz,
Phys.\ Rev.\ D {\bf 89}, 10, 103537 (2014)
[arXiv:1310.8133 [hep-th]].

\bibitem{Ijjas:2014fja}
A.~Ijjas, J.~L.~Lehners and P.~J.~Steinhardt,
Phys.\ Rev.\ D {\bf 89}, 12, 123520 (2014)
[arXiv:1404.1265 [astro-ph.CO]].



\bibitem{Cai:2015yza}
Y.~Cai, Y.~T.~Wang and Y.~S.~Piao,
Phys.\ Rev.\ D {\bf 93}, 6, 063005 (2016)
[arXiv:1510.08716 [astro-ph.CO]].

\bibitem{Cai:2015dta}
Y.~Cai, Y.~T.~Wang and Y.~S.~Piao,
Phys.\ Rev.\ D {\bf 91}, 103001 (2015)
[arXiv:1501.06345 [astro-ph.CO]].

\bibitem{Cai:2015ipa}
Y.~Cai, Y.~T.~Wang and Y.~S.~Piao,
JHEP {\bf 1602} (2016) 059
[arXiv:1508.07114 [hep-th]].



\bibitem{Nakashima:2010sa}
M.~Nakashima, R.~Saito, Y.~i.~Takamizu and J.~Yokoyama,
Prog.\ Theor.\ Phys.\  {\bf 125}, 1035 (2011)
[arXiv:1009.4394 [astro-ph.CO]].


\bibitem{Park:2012rh}
M.~Park and L.~Sorbo,
Phys.\ Rev.\ D {\bf 85}, 083520 (2012)
[arXiv:1201.2903 [astro-ph.CO]].

\bibitem{Bartolo:2013exa}
N.~Bartolo, D.~Cannone and S.~Matarrese,
JCAP {\bf 1310}, 038 (2013)
[arXiv:1307.3483 [astro-ph.CO]].

\bibitem{Achucarro:2014msa}
A.~Achucarro, V.~Atal, B.~Hu, P.~Ortiz and J.~Torrado,
Phys.\ Rev.\ D {\bf 90} (2014) 2,  023511
[arXiv:1404.7522 [astro-ph.CO]].


\bibitem{Saito:2013aqa} 
R.~Saito and Y.~i.~Takamizu,
JCAP {\bf 1306}, 031 (2013)
[arXiv:1303.3839 [astro-ph.CO]].


\bibitem{Mizuno:2014jja} 
S.~Mizuno, R.~Saito and D.~Langlois,
JCAP {\bf 1411}, no. 11, 032 (2014)
[arXiv:1405.4257 [hep-th]].













\bibitem{Banerjee:2016hom}
  S.~Banerjee and E.~N.~Saridakis,
  Phys.\ Rev.\ D {\bf 95}, 6, 063523 (2017)
  [arXiv:1604.06932 [gr-qc]].

\bibitem{Biswas:2011ar}
  T.~Biswas, E.~Gerwick, T.~Koivisto and A.~Mazumdar,
  Phys.\ Rev.\ Lett.\  {\bf 108}, 031101 (2012)
  [arXiv:1110.5249 [gr-qc]];
  T.~Biswas, A.~S.~Koshelev, A.~Mazumdar and S.~Y.~Vernov,
  JCAP {\bf 1208}, 024 (2012)
  [arXiv:1206.6374 [astro-ph.CO]].

\bibitem{Vasilic:2017myy}
M.~Vasilic,
Phys.\ Rev.\ D {\bf 95}, 12, 123506 (2017)
[arXiv:1704.02589 [gr-qc]].

\bibitem{Li:2016xjb}
Y.~B.~Li, J.~Quintin, D.~G.~Wang and Y.~F.~Cai,
JCAP {\bf 1703}, 03, 031 (2017)
[arXiv:1612.02036 [hep-th]].

\bibitem{Odintsov:2015zza}
S.~D.~Odintsov and V.~K.~Oikonomou,
Phys.\ Rev.\ D {\bf 92} (2015) 2,  024016
[arXiv:1504.06866 [gr-qc]];
S.~D.~Odintsov and V.~K.~Oikonomou,
Phys.\ Rev.\ D {\bf 91} (2015) 6,  064036
[arXiv:1502.06125 [gr-qc]];
S.~D.~Odintsov and V.~K.~Oikonomou,
Phys.\ Rev.\ D {\bf 90} (2014) 12,  124083
[arXiv:1410.8183 [gr-qc]].

\bibitem{Hendi:2016tiy}
S.~H.~Hendi, M.~Momennia, B.~Eslam Panah and M.~Faizal,
Astrophys.\ J.\  {\bf 827}, 2, 153 (2016)
[arXiv:1703.00480 [gr-qc]];
S.~H.~Hendi, M.~Momennia, B.~Eslam Panah and S.~Panahiyan,
Phys. Dark Universe 16, 26 (2017)
[arXiv:1705.01099 [gr-qc]].


\bibitem{Giovannini:2016jkf}
M.~Giovannini,
Phys.\ Rev.\ D {\bf 95}, 8, 083506 (2017)
[arXiv:1612.00346 [hep-th]].




\bibitem{Easson:2011zy}
D.~A.~Easson, I.~Sawicki and A.~Vikman,
JCAP {\bf 1111}, 021 (2011)
[arXiv:1109.1047 [hep-th]].





\end{thebibliography}
 \end{document}